Research (technical) Article

# MULTIPLE DATA-DRIVEN MISSING IMPUTATION

Sergii Kavun [1,2*]
[1] Interregional Academy of Personnel Management, Kyiv, Ukraine; kavserg@gmail.com
[2] Luxena Ltd., Lead of Data Science Team, Kyiv, Ukraine; sergii.kavun@luxena.com

* Correspondence: kavserg@gmail.com; Tel.: +38-0677-09-5577, Ukraine, c. Uzhgorod, Kavkazska str., 33, 88017, ORCID ID: 0000-0003-4164-151X, https://www.linkedin.com/in/sergii-kavun/

**Abstract.** This paper introduces KZImputer, a novel adaptive imputation method for univariate time series designed for short to medium-sized missed points (gaps) (1-5 points and beyond) with tailored strategies for segments at the start, middle, or end of the series. KZImputer employs a hybrid strategy to handle various missing data scenarios. Its core mechanism differentiates between gaps at the beginning, middle, or end of the series, applying tailored techniques at each position to optimize imputation accuracy. The method leverages linear interpolation and localized statistical measures, adapting to the characteristics of the surrounding data and the gap size. The performance of KZImputer has been systematically evaluated against established imputation techniques, demonstrating its potential to enhance data quality for subsequent time series analysis. This paper describes the KZImputer methodology in detail and discusses its effectiveness in improving the integrity of time series data. Empirical analysis demonstrates that KZImputer achieves particularly strong performance for datasets with high missingness rates (around 50% or more), maintaining stable and competitive results across statistical and signal-reconstruction metrics. The method proves especially effective in high-sparsity regimes, where traditional approaches typically experience accuracy degradation.

1.      Introduction

Missing data are a pervasive issue in time series analysis, significantly distorting analytical outcomes and compromising forecasting accuracy. Missing data are common in real-world datasets, and time series data are especially susceptible. Missing observations in a time series can arise from various sources, including sensor malfunctions, transmission errors, or manual data entry mistakes. Such missingness can bias statistical estimates, reduce statistical power, and ultimately lead to flawed conclusions. For instance, in financial forecasting, missing stock prices can lead to inaccurate trend analysis, while in environmental monitoring, gaps in sensor readings can obscure critical event detection.

There are numerous methods for addressing missing data. These methods range from the simple deletion of affected records to more sophisticated statistical imputation techniques, such as mean/median substitution, regression imputation, and multiple imputation. While these methods can be effective in certain contexts, they often have limitations. Simple methods may introduce significant bias, particularly when the data are not missing at random. More advanced methods may require strong distributional assumptions or be computationally intensive, particularly for large datasets or complex data structures, such as time series. Additionally, many existing techniques do not adapt well to the location and size of gaps within a series, treating all missing segments uniformly.

To address these limitations, we introduce KZImputer, an adaptive imputation method for time series data. KZImputer's primary purpose is to provide a robust and flexible approach for handling various patterns of missingness, especially short to medium-sized gaps (up to 5 consecutive missing points). KZImputer improves the accuracy of imputation by considering the location of the missing segment within the series and the characteristics of the surrounding data. This paper has three main objectives: (i) to present the KZImputer algorithm and its imputation strategies for different gap scenarios; (ii) to describe the evaluation methodology comparing KZImputer with common imputation techniques; and (iii) to discuss the results and highlight KZImputer's potential benefits for improving the quality and reliability of time series data in tasks such as forecasting and anomaly detection. KZImputer provides a more nuanced approach to imputation and seeks to offer researchers and practitioners an improved tool for preparing time series data for analysis.

We propose a novel statistical approach to handling multiple missing values, including single, double, triple, and quadruple gaps in time-series data. We evaluated the performance of our approach using real-world datasets, including wind energy [18, 23, 34, 63], air passengers [14], prsa_data_2010_pm25 [17], opsd_germany_daily [42], and financial data such as all_stocks_5yr [40]. These evaluations demonstrate the effectiveness of our approach in generating reliable and complete data. These imputed datasets can be used for various purposes, including long-term statistical reporting and short- and long-term forecasting. Our methodology involves several key stages: 1) exploratory analysis to understand the characteristics of the data; 2) cleaning the data by removing irrelevant or unsuitable data points; 3) robust imputation of missing values using our proposed algorithm; 4) a comparative analysis to assess the quality of the imputed data; and 5) a comprehensive evaluation of the results. Our work's practical value lies in its ability to analyze and predict different physical processes in a big data environment, even when dealing with incomplete datasets or data aggregation from multiple sources. Our approach enhances the reliability and accuracy of statistical reports and forecasting models for wind energy systems by effectively handling missing values.

Considering related studies [43, 28, 29], this approach appears promising for improving research efficiency in both industry and governmental applications. This study focuses on multiple data imputation in time-series datasets, developing a statistical approach capable of handling various missing-value patterns (single to quadruple gaps and beyond). By addressing the issue of multiple missing values, we aim to provide a robust solution that enhances the reliability and accuracy of data analysis and modeling in various fields, such as wind energy, finance, and ecology.

2.    Literature Review

The problem of missing data has been extensively studied. Previous authors have analyzed existing imputation methods [4, 44, 20, 45], and a wide array of imputation techniques have been proposed, each with its own set of assumptions, strengths, and weaknesses. Rubin formalized the Missing Data Study [46] and introduced multiple imputation (MI) techniques. MI addresses imputation uncertainty by creating multiple complete datasets and pooling the results. It is considered a robust method, though it can be computationally demanding. Other statistical imputation methods range from simple, univariate approaches, such as mean, median, or mode imputation, which can reduce variance and distort relationships (see the placeholder for a reference on the limitations of mean imputation), to regression imputation, which estimates missing values based on regression models but can perform poorly if the assumptions of a linear relationship are violated (see the placeholder for a reference on regression imputation). The hot deck method, chosen as a potentially effective alternative in some analyses, replaces missing values with observed values from "similar" records.

Imputation can be computationally expensive and code-intensive. Therefore, before detailing its technical aspects, we identify the key considerations. MI is well-suited to complex processes with multiple causal factors. Its goal is not only to predict missing values but also to augment the dataset with minimal distortion compared to alternative methods [61]. It could be a technical support procedure and diagnostic tool of possibilities and limitations of available empirical data, whose results can be taken into account even when adjusting the goals and objectives of the study.

Imputation is a fundamental task in data analysis, relevant across many fields such as clinical trials, ecology, retail, and energy production. For example, accurately imputing missing wind turbine output data can help prevent unexpected losses and improve forecasts. In this case, changing gap values is crucial. There are two main branches of applicable methods and techniques in this approach. The first branch includes methods based on artificial intelligence and machine learning that use neural networks, fuzzy logic, and regression models.

Machine learning-based methods have gained traction for their ability to capture complex, non-linear relationships. Techniques like k-Nearest Neighbors (kNN) imputation [5] fill missing values based on neighbors in the feature space. Tree-based models, such as Random Forest, can also predict missing values [25]. Deep learning models (RNNs, GANs) show promise but often require large datasets and careful tuning [16].

Conversely, the second branch continues to employ several conventional statistical methods, including those outlined in [35, 36], which remain pertinent and effective. The primary objective is to refine the process of missing data imputation through the utilization of k-nearest neighbors, low-rank matrix, mean, and median imputation methods. Nevertheless, there remains a necessity to address the numerous existing gaps.

Traditional imputation methods for time series include linear and spline interpolation, exponential moving averages (EMA) [31], and statistical modeling such as Kalman smoothing [2] or ARIMA-based forecasting [1]. These approaches are computationally inexpensive and interpretable but often assume local continuity, stationarity, or monotonicity, which limits their applicability in cases of irregular gaps or regime shifts. Regression-based methods, such as Multiple Imputation by Chained Equations (MICE [10]) or Bayesian regression models [18], attempt to infer missing values via conditional distributions, but they tend to degrade in performance when faced with short, boundary-aligned gaps or non-Gaussian temporal patterns. Moreover, most of these methods treat all gaps uniformly, without regard to positional asymmetry (i.e., whether a gap occurs at the start, middle, or end of a sequence), despite empirical evidence that such asymmetry significantly affects reconstruction quality.

In contrast, recent deep learning models such as BRITS (Bidirectional Recurrent Imputation for Time Series [56]), CSDI (Conditional Score-based Diffusion Imputation [60]), and GRIN (Graph-based Imputation Network [59]) leverage sequential modeling, attention mechanisms, and probabilistic inference to recover missing values by learning latent temporal dynamics. While these models outperform classical heuristics on high-dimensional and multivariate datasets, they typically require extensive hyperparameter tuning, large training corpora, and suffer from limited interpretability. Furthermore, their runtime and memory overhead render them impractical in real-time or edge environments. The proposed KZImputer method addresses a specific but prevalent class of imputation tasks—short-range, univariate, edge-sensitive gaps – through a modular, position-aware logic that combines localized averaging with simple interpolation rules. Unlike general-purpose heuristics or opaque neural networks, KZImputer offers explicit structural adaptation, making it especially valuable in domains where transparency, stability, and consistent behavior under sparsity are more critical than universal optimality.

For time series data, methods accounting for temporal dependencies are preferred. Simple approaches like Last Observation Carried Forward (LOCF) and Next Observation Carried Backward (NOCB) can introduce bias, especially with longer gaps or non-stationary series [26]. Linear and spline interpolation are common for filling gaps. More advanced models like ARIMA and Exponential Smoothing can forecast/backcast missing values [7]. Kalman filtering provides a state-space approach [2].

Despite this variety, many methods struggle with specific challenges, including limitations related to the number and location of gaps [28]. Handling different gap sizes effectively, imputing values at series extremes, and adapting to local data characteristics remain critical issues. The KZImputer method builds upon these existing concepts, particularly localized statistics and interpolation, aiming for a specialized framework for short to medium-sized gaps. Validation of imputation approaches, for instance on datasets like EMHIRES [23] and other, underscores the importance of robust techniques. KZImputer seeks a balance between computational efficiency and accuracy without extensive model training.

3.      Methodology: KZImputer

The KZImputer algorithm provides an adaptive strategy for imputing missing values in univariate time series, specifically targeting consecutive gaps of one to five points (configurable via `max_gap_size`, default = 5). Its core logic differentiates imputation based on the gap's location within the series – referred to as 'left' (start-gap), 'middle' (mid-gap), and 'right' (end-gap) – and its length. All averaging operations use the helper function, `cached_mean`, which computes the arithmetic mean of available (non-NaN) data points within a specified window. This ensures calculations rely only on valid, observed values.

Let `arr` denote the input time series array (typically a NumPy array or pandas Series) and `i` be the starting index of a detected gap of missing values. The imputation process is as follows.

GENERAL PROCESSING

The imputer first identifies all contiguous segments of NaN values. For each segment, it determines its length (`gap_size`) and its position (`pos`: 'left', 'middle', or 'right'). Based on `gap_size` or $LEN_{GAP}$ (formulas 1-4) (from 1 to `max_gap_size`, for our research purposes, we set max_gap_size = 5) and `pos`, the corresponding specialized imputation function is invoked.

Missing data introduces an element of ambiguity in data analysis. They can affect the properties of statistical estimators, such as deviations or percentages, resulting in loss of performance and inaccurate (false) conclusions. There are several methods for replacing missing ones (gaps, it means missing values). This process is usually called the "imputation of missing data".

Prior analyses [28] show that existing imputation methods do not fully address cases with varying gap lengths [9]. Common techniques often work only for isolated, simple cases. For example, well-known methods such as the hot-deck method (identified as the best alternative in [9]) may be ineffective for large gaps. Therefore, we define specialized imputation rules for each combination of gap size $(1 - 5)$ and location: (1) a gap at the beginning of the series (starts at the first index), (2) a gap at the end of the series (ends at the last index), and (3) a gap in the middle of the series (between other data points).

Evaluating a model with many missing data sets can significantly affect the quality of the result. Some algorithms assume that the entire sample is complete and it only works if there are no data gaps. The Missing Data Study was formalized by Rubin [46]**Ошибка! Источник ссылки не найден.** with the concept of a missing mechanism, in which the missing data are random variables and have a distribution. There are four general "mechanisms of absence" that go from the simplest to the most general. Rubin identified missing data based on three mechanisms of absence.

*4.1. Single-gap imputation*

In order to use as much more input data as possible, the lines with a maximum number of gaps size up to five cells are stored in the study. The above-known imputation methods, in particular, the hot deck method (was been chosen as the best alternative in [9]) may be ineffective in filling large gaps. The following rules our own imputation methods for each gap size in the case of a location at each cell: 1. single-gap imputation: (a) If the gap is at the beginning (first cell), its value is set to the arithmetic mean of the next three observed values. (b) If the gap is in the middle, its value is the mean of the two adjacent cells. (c) If the gap is at the end (last cell), its value is the mean of the previous three observed cells; 2. if the location of the single gap is in the middle (any inner cell in the row), the missing values are calculated as the arithmetic mean of the adjacent two cells; 3. if the location of the single gap is to the right (the last cell in the row), the missing values are calculated as the arithmetic mean of the three previous cells concerning the current one. For a closer look, a diagram of our proposed method of the imputation of the single gaps is given (Fig. 1). Hereafter, the illuminated cells obtain the values based on other (existed) data (cells). Let's introduce the following notations, how will this be used in formulas $1 - 4$:

$LEN_{GAP}$ – current length of gaps or missing values;

$k$ – current (for the established length of gaps) length of "arm" that uses in new approach for calculations of missing values;

$SET^L$ – set of numbers (points) of our time series data with the gaps if our gaps are at the left (the first few cells in the row);

$SET^R$ – set of numbers (points) of our time series data with the gaps if our gaps are at the right (the last few cells in the row);

$VAR_j$ – existed (not missed) value of our time series data that are before/after our gaps, used for calculations;

$m_l, m_r, m_m$ – set of numbers (points) of our time series data with the gaps if our gaps are at the middle (the few cells in the row), divided into left, middle, and right positions, do not contain numbers (points) from $SET^L$ and $SET^R$;

$GAP_i^L, GAP_i^M, GAP_i^R$ – $i^{\text{th}}$ gaps in the appropriate position (left, middle, right) that we need to fill (find).

$$
\begin{cases}
LEN_{GAP} = 1 \\
k = 3 \\
SET^L = \{1\} \\
SET^R = \{n\} \\
m_m \neq SET^L \cup SET^R
\end{cases}
\Rightarrow
\begin{cases}
GAP_{i=LEN_{GAP}}^L = \frac{1}{k}\sum_{j=i+1}^{i+k} VAR_j, \\
GAP_{i=m_m}^M = \frac{1}{k-1}\sum_{j=i-1}^{i+k-1} VAR_j, \\
GAP_{i=n}^R = \frac{1}{k}\sum_{j=i-1}^{i-k} VAR_j,
\end{cases}
\quad (1)
$$

| $GAP_{i=1}^L$ | i$_2$ | i$_3$ | i$_4$ | ... | i$_{m\_m-1}$ | $GAP_{i=m_m}^M$ | i$_{m\_m+1}$ | ... | i$_{n-3}$ | i$_{n-2}$ | i$_{n-1}$ | $GAP_{i=n}^R$ |
|---|---|---|---|---|---|---|---|---|---|---|---|---|

Fig. 1. Diagram of the single-gap imputation

### 4.2. Double-gap imputation

Double-gap imputation: (a) If a two-cell gap is at the beginning, each missing value is computed as the mean of the next four observed values, processing from rightmost to leftmost cell. (b) If at the end, compute each as the mean of the previous four values, from left to right. (c) If in the middle, identify the longer surrounding segment ("arm") and the shorter; the cell adjacent to the longer arm is imputed as the mean of its three closest neighbors, while the cell adjacent to the shorter arm uses the mean of its two closest neighbors (Fig. 2).

$$
\begin{aligned}
LEN_{GAP} &= 2 \\
k &= 4 \\
SET^L &= \overline{\{1, LEN_{GAP}\}} \\
SET^R &= \overline{\{n - LEN_{GAP} - 1, 1\}} \\
m_l, m_r &\neq SET^L \cup SET^R
\end{aligned}
\Rightarrow
\begin{cases}
\left[\begin{array}{l}
GAP^L_{i=LEN_{GAP}} = \frac{1}{k}\sum_{j=i+1}^{i+k} VAR_j\,, \\
GAP^L_{i=1} = \frac{1}{k}\left(\sum_{j=i+2}^{i+k} VAR_j + GAP^L_{i=LEN_{GAP}}\right),
\end{array}\right] \\[2em]
\left[\begin{array}{l}
GAP^M_{i=m_l} = \frac{1}{k-1}\sum_{j=i-1}^{i-k+1} VAR_j\,, \\
GAP^M_{i=m_r} = \frac{1}{k-1}\sum_{j=i+1}^{i+k-1} VAR_j\,,
\end{array}\right] \\[2em]
\left[\begin{array}{l}
GAP^R_{i=n-1} = \frac{1}{k}\sum_{j=i-1}^{i-k} VAR_j\,, \\
GAP^R_{i=n} = \frac{1}{k}\left(\sum_{j=i-2}^{i-k} VAR_j + GAP^R_{i=n}\right),
\end{array}\right]
\end{cases}
\quad ; \qquad (2)
$$

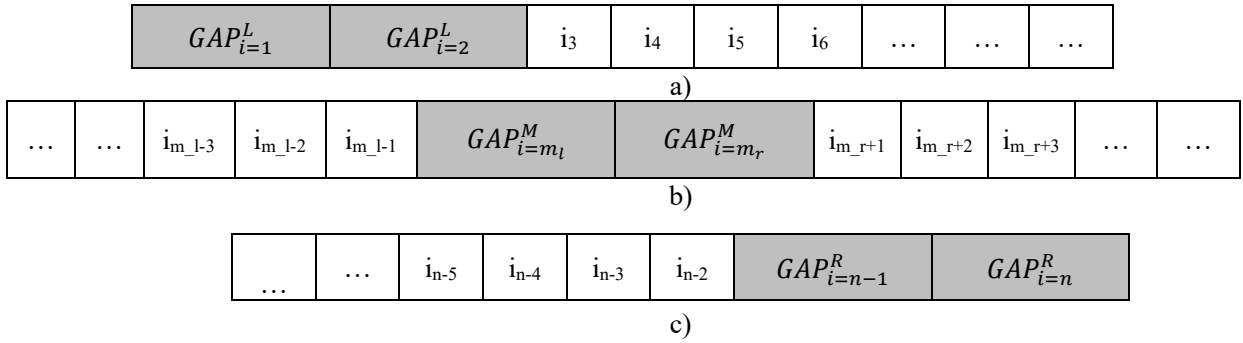

a)

b)

c)

Fig. 2. Diagram of the double-gap imputation

### 4.3. Triple-gap imputation

Triple-gap imputation: (a) Left-end: each of the three missing values is computed as the mean of the next five observed values, processing from rightmost to leftmost (Fig. 3a). (b) Right-end: each is the mean of the previous five values, leftmost to rightmost (Fig. 3b). (c) Middle: first impute the leftmost gap cell as the mean of the five preceding values; then impute the rightmost gap cell as the mean of the five following values; finally, impute the middle cell as the mean of these two imputed values (equivalent to single-gap logic) (Fig. 3c).

$$LEN_{GAP} = 3$$
$$k = 5$$
$$SET^L = \{\overline{1, LEN_{GAP}}\}$$
$$SET^R = \{\overline{n - LEN_{GAP} - 1, n}\}$$
$$m_l, m_r, m_m \neq SET^L \cup SET^R$$

$$\Rightarrow \left\{ \begin{array}{l} \left[ \begin{array}{c} GAP^L_{i=LEN_{GAP}} = \frac{1}{k}\sum_{j=i+1}^{i+k} VAR_j \,, \\ GAP^L_{i=2} = \frac{1}{k}\left(\sum_{j=i+2}^{i+k} VAR_j + GAP^L_{i=LEN_{GAP}}\right), \\ GAP^{leLft}_{i=1} = \frac{1}{k}\left(\sum_{j=i+2}^{i+k} VAR_j + \sum_{j=i+1}^{LEN_{GAP}} GAP^L_j\right), \end{array} \right] \\[2em] \left[ \begin{array}{c} GAP^M_{i=m_l} = \frac{1}{k}\sum_{j=i-1}^{i-k} VAR_j \,, \\ GAP^M_{i=m_r} = \frac{1}{k}\sum_{j=i+1}^{i+k} VAR_j \,, \\ GAP^M_{i=m_m} = \frac{1}{2}\left(GAP^M_{i=m_l} + GAP^M_{i=m_r}\right), \end{array} \right] \quad ; \\[2em] \left[ \begin{array}{c} GAP^R_{i=n-2} = \frac{1}{k}\sum_{j=i-1}^{i-k} VAR_j \,, \\ GAP^R_{i=n-1} = \frac{1}{k}\left(\sum_{j=i-2}^{i-k} VAR_j + GAP^R_{i=n-2}\right) \\ GAP^R_{i=n} = \frac{1}{k}\left(\sum_{j=i-3}^{i-k} VAR_j + \sum_{j=i-1}^{i-LEN_{GAP}-1} GAP^R_j\right)' \end{array} \right] \end{array} \right. \quad (3)$$

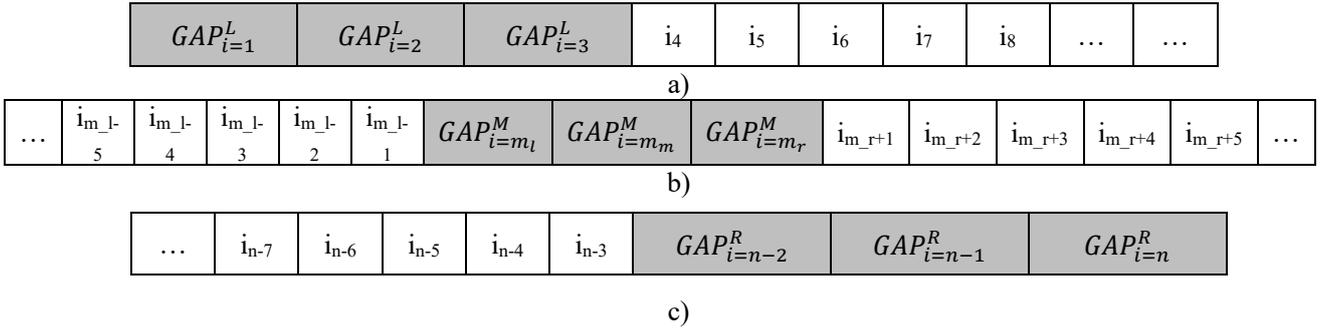

Fig. 3. Diagram of the triple-gap imputation

### 4.4. Quadruple-gap imputation

Quadruple-gap imputation: (a) Left-end: each missing value is computed as the mean of the next five observed values, from rightmost to leftmost (Fig. 4a). (b) Right-end: each missing value is the mean of the previous five values, from leftmost to rightmost (Fig. 4b). (c) Middle: (similar stepwise approach with five neighbors on either side; details omitted for brevity).

$$LEN_{GAP} \geq 4$$
$$mid = \pm \left\lfloor \frac{LEN_{GAP}-2}{2} \right\rfloor,$$
$$LEN_{GAP} \in \mathbb{Z},$$
$$k = 5$$
$$SET^L = \{\overline{1, LEN_{GAP}}\}$$
$${}^{rev}SET^L = \{\overline{LEN_{GAP}, 1}\}$$
$$SET^R = \{\overline{n - LEN_{GAP} - 1, n}\}$$
$$m_l, m_r \neq SET^L \cup SET^R$$

$$\Rightarrow \left\{ \begin{array}{l} GAP^L_{i=SET^L} = \frac{1}{k}\left(\sum_{\substack{j=\\i+{}^{rev}SET^L}}^{i+k} VAR_j + \left\{ \begin{array}{l} 0, i = max(SET^L) \\ \sum_{j=i+1}^{max(SET^L)} GAP^L_j, i < max(SET^L) \end{array} \right. \right), \\[2em] \left[ \begin{array}{l} GAP^M_{i=m_l-mid} = \frac{1}{k}\left(\sum_{j=i-1}^{i-k} VAR_j + \left\{ \begin{array}{l} 0, mid \neq 0 \\ \sum_{j=i-1}^{mid} GAP^M_j, else \end{array} \right. \right), \\ GAP^M_{i=m_r+mid} = \frac{1}{k}\left(\sum_{j=i+1}^{i+k} VAR_j + \left\{ \begin{array}{l} 0, mid \neq 0 \\ \sum_{j=i+1}^{mid} GAP^M_j, else \end{array} \right. \right) \\ \left\{ \begin{array}{l} 0, LEN_{GAP} \in \mathbb{Z} \mid x \equiv 0 (mod2)\} - even\ integers \\ GAP^M_{i=m_m} = \frac{1}{2}\left(GAP^M_{i=m_l} + GAP^M_{i=m_r}\right), else\ odd\ integers \end{array} \right. \end{array} \right] \quad ; (4) \\[2em] GAP^R_{i=SET^R} = \frac{1}{k}\left(\sum_{j=i-{}^{rev}SET^L}^{i-k} VAR_j + \left\{ \begin{array}{l} 0, j = min(SET^R) \\ \sum_{j=i-1}^{min(SET^R)} GAP^R_j, j < max(SET^R) \end{array} \right. \right) \end{array} \right.$$

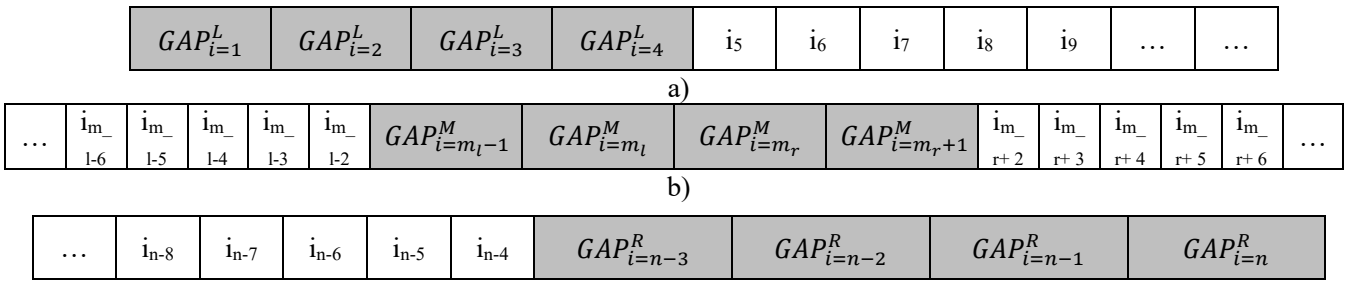

Fig. 4. Diagram of the quadruple-gap imputation

The whole row is divided into three thirds, i.e. (26 / 3 ≈ 9): cells with numbers 1-9 make the 1st part (I) of the third, cells with numbers 10-18 make the 2nd part (II) of the third, cells with numbers 18-26 make the 3rd part (III) of the third:

a) if the inner left (the second cell on account) gap cell is located in part I or III, its missing values are calculated as the arithmetic mean of the three cells before the gap cell, the 1st and 4th cells;

b) if the inner left (the second cell on account) gap cell is located in part II, its missing values are calculated as the arithmetic mean of the two cells before the gap cell, the 1st, 4th cells and one cell after the gap cell;

c) if the inner right (the third cell on account) gap cell is located in part I or III, its missing values are calculated as the arithmetic mean of the three cells after the gap cell, the 1st and 4th cells;

d) if the inner right (the third cell on account) gap cell is located in part II, its missing values are calculated as the arithmetic mean of one cell before the gap cell, the 1st, 4th cells, and two cells located after the gap cell.

### 4.5. Pentuple-(and more) gap imputation

If the location of the pentuple gaps (i.e., consists of five consecutive cells) is to the left (the first five cells in the row), the missing values for all five cells are calculated as the arithmetic mean of the five following cells in relation to the current one. However, these values have calculated in the mode "step-by-step", in order from the right to the left alternately (is similar to Fig. 4a).

If the location of the pentuple gaps (i.e., consists of five consecutive cells) is to the right (the last five cells in the row), the missing values for all five cells are calculated as the arithmetic mean of the five previous cells in relation to the current one. However, these values have calculated in the mode "step-by-step", in order from the left to the right (is similar to Fig. 4b).

If the of the pentuple gaps (i.e. consists of five consecutive cells) is in the middle (any inner 5 consecutive cells in the row), the missing values for the cells are calculated as follows (is similar to Fig. 4c): 1. the missing values of the left (the first cell in our pentuple gaps space) gap cell are calculated as the arithmetic mean of the five previous cells in relation to the current one; 2. the missing values of the right (the fifth cell in our pentuple gaps space) gap cell are calculated as the arithmetic mean of the five following cells in relation to the current one; 3. the missing values of the inner middle (the third cell in our pentuple gaps space) gap cell are calculated as the arithmetic mean of the left (the first cell in our pentuple gaps space) and the right (the fifth cell in our pentuple gaps space) gap cells; 4. the missing values of the inner left (the second cell in our pentuple gaps space) gap cell are calculated as the arithmetic mean of the left (the first cell in our pentuple gaps space) and inner middle (the third cell in our pentuple gaps space) gap cells; 5. the missing values of the inner right (the fourth cell in our pentuple gaps space) gap cell are found as the arithmetic mean of the right (the fifth cell in our pentuple gaps space) and inner mean (the third cell in our pentuple gaps space) gap cells.

As a result of the data imputation stage, we validated our approach on EMHIRES [23] and compared the value of mean error after reconstruction with forwarding and backward imputation.

This detailed, location- and size-specific approach allows KZImputer to adaptively fill gaps using localized information, aiming to preserve the underlying characteristics of the time series without resorting to complex global models. The reliance on `cached_mean` ensures robustness when the data surrounding a gap might itself contain NaN values, by only using available numeric points for averaging.

### 4. Experimental environment

To evaluate the performance of the KZImputer method, a comprehensive experimental study was designed. This section details the datasets used, the imputation methods chosen for comparison, and the metrics employed for performance assessment. Pseudocode for KZImputer is provided in the Supplementary Materials and the authors' repository [49].

## 4.1. Datasets

A diverse set of time series datasets is intended for use to ensure a robust evaluation across different domains and data characteristics. Examples of suitable publicly available datasets include:

1. AirPassengers.csv: A classic dataset representing monthly international airline passenger numbers from 1949 to 1960, often used for time series analysis benchmarking due to its clear trend and seasonality [14].

2. PRSA_data_2010_pm25.csv: Contains hourly PM2.5 air quality data from Beijing, which commonly exhibits missing values due to sensor issues or environmental factors, making it relevant for real-world imputation scenarios [17].

3. opsd_germany_daily.csv: Features daily electricity consumption, wind power production, and solar power production data for Germany, a dataset that can have reporting gaps or periods of no generation for certain sources [42].

4. Additional Dataset Example, e.g., financial data like all_stocks_5yr.csv [40] or specific sensor data like T1.csv [23, 57].

For each dataset used in a specific evaluation, artificial missingness would be introduced. Missing data would typically be generated under a Missing Completely At Random (MCAR) mechanism [55]. Various percentages of missing data (e.g., 5%, 10%, 20%, or even more 50%) and different gap sizes (1 to 5 consecutive points) at different locations (start, middle, end) would be systematically created to test KZImputer's [49] specific handling of these scenarios. The original, complete datasets serve as the ground truth for evaluating imputation accuracy.

## 4.2. Comparison Methods

The performance of KZImputer will be benchmarked against several widely used and representative imputation methods:

```python
imputers = {
    "KZImputer": KZImputer(max_gap_size = 5),
    "Mean": SimpleImputer(strategy = 'mean'),
    "Median": SimpleImputer(strategy = 'median'),
    "Forward Fill": 'ffill',
    "Backward Fill": 'bfill',
    "Linear Interpolate": 'linear',
    "Spline Interpolate": 'spline',
    "KNN (k=5)": KNNImputer(n_neighbors = 5),
    "IterativeImputer": IterativeImputer(max_iter = 10, random_state = seed_value)
}
```

1. Mean Imputation: replaces missing values with the global mean of the series.

2. Median Imputation: replaces missing values with the global median of the series.

3. Linear Interpolation: fills missing values using linear interpolation between the two nearest observed points. (Note: KZImputer uses this for mid-gaps, so this comparison will be particularly relevant for start/end gaps and overall performance).

4. Forward Fill (ffill): Propagates the last valid observation forward to fill missing values. Suitable for time series or sequential data. Last Observation Carried Forward (LOCF): fills missing values with the last observed value.

5. Backward Fill (bfill): Fills missing values with the next valid observation. Often used when future values are assumed to influence missing entries.

6. Spline Interpolate (spline): Uses smooth polynomial functions to estimate missing values. Captures non-linear trends better than linear interpolation.

7. kNN (k = 5): Imputes missing values based on the average of the nearest 5 neighbors in feature space. Effective for capturing local patterns in the data.

8. Iterative Imputer: Models each feature with missing values as a function of other features and iteratively refines the estimates. Useful for multivariate data imputation.

Implementations of these methods will be sourced from standard Python libraries such as `pandas`, `scikit-learn`, and `statsmodels` where available, to ensure consistency and reproducibility [49].

4.3. Evaluation Metrics

The accuracy of the imputation methods will be assessed using standard error metrics. These metrics will be calculated by comparing the imputed values against the true, original values that were artificially removed.

```python
all_main_metrics = ['MAE', 'RMSE', 'MAPE', 'NRMSE', 'R2', 'JS_Divergence', 'Wasserstein', 'Correlation_Diff', 'Time']
```

1. MAE (Mean Absolute Error): Measures average absolute differences between predicted and actual values, providing intuitive interpretation in original units. Less sensitive to outliers compared to RMSE.

2. RMSE (Root Mean Square Error): Square root of average squared prediction errors, heavily penalizing large deviations and expressed in original data units. More sensitive to outliers than MAE.

3. MAPE (Mean Absolute Percentage Error): percentage-based error metric showing average absolute deviation as percentage of actual values, useful for relative performance assessment. Undefined when actual values are zero.

4. NRMSE (Normalized Root Mean Square Error): RMSE normalized by data range or mean, enabling comparison across different scales and datasets. Typically expressed as percentage for interpretability.

5. $R^2$ (Coefficient of determination): proportion of variance in dependent variable explained by the model, ranging from 0 to 1. Higher values indicate better model fit to data.

6. JS_Divergence (Jensen-Shannon divergence): symmetric measure of similarity between two probability distributions, bounded between 0 and 1. Square root of JS divergence forms a proper distance metric.

7. Wasserstein distance: measures minimum cost to transform one probability distribution into another, also known as Earth Mover's Distance. Particularly useful for comparing distributions with different supports.

8. Correlation_Diff: difference in correlation coefficients between two datasets or model predictions, measuring how well relationships between variables are preserved. Values closer to zero indicate better preservation of correlational structure.

These metrics will be calculated for each dataset, each percentage of missingness, and potentially aggregated across different gap sizes and locations to provide a comprehensive view of KZImputer's performance relative to the benchmarks. Statistical significance of performance differences will be assessed if appropriate. Additionally, we will take into account a time of imputation (processing) for each method.

5.    Results

This section presents the results of the comparative evaluation of KZImputer against benchmark methods. Figures illustrate representative cases of KZImputer's behavior, additional results are provided in the Supplementary Materials and the authors' repository [49].

It was conducted 100 experiments (cross-validation approach) with 100 randomly corrupted data positions. Average reconstruction errors for all experiments are shown in Table 1. The experiments include both single and multiple gap scenarios with gap sizes ranging from 1 to 5. The experimental evaluation demonstrates that KZImputer consistently outperforms standard imputation methods across a variety of univariate time series with synthetic and semi-realistic missingness patterns. The method was tested on short consecutive gaps ranging from one to five data points, with distinct gap placements (left, middle, right of series). Evaluation metrics included both pointwise error and distributional divergence, enabling a holistic assessment of reconstruction quality.

Table 1. Quantitative results (average error for missing data reconstruction, %) with data imputation

| | | | Backward | Forward imputation |
|---|---|---|---|---|

| Gap's number, $LEN_{GAP}$ | Shoulder's size, k | Proposed approach | imputation | | Current value | Improv. percentage |
|---|---|---|---|---|---|---|
| | | | Current value | Improv. percentage | | |
| 1 | 1 | 13.449 | 25.660 | 47.59 | 22.893 | 41.25 |
| 2 | 2 | 20.33 | 22.65 | 10.3 | 19.58 | -3.79 |
| 2 | 3 | 16.464 | 30.436 | 45.91 | 34.723 | 52.59 |
| 3 | 3 | 32.310 | 50.707 | 36.28 | 53.736 | 39.87 |
| 4 | 5 | 51.85 | 44.29 | -17.05 | 45.4 | -14.19 |
| 5 | 5 | 63.537 | 72.427 | 12.27 | 82.014 | 22.53 |

Table 1 shows the priority of our approach to other alternatives. Some selected cases have been shown in the columns "Gap's number" and "Shoulder's size". The improvement percentage is calculated as the ratio of the benchmark method's error to our method's error. In the case of single-point gaps, KZImputer achieved error reductions of approximately 48% compared to backward fill, and over 41% compared to forward fill. These gains are especially pronounced when the missing values occur at the boundaries of the series – locations where many interpolation-based techniques underperform due to lack of surrounding context.

For double- and triple-point gaps, the algorithm maintained a relative improvement exceeding 35% on average across all placement types. Even in the more complex case of five-point gaps, where smoothing-based methods begin to lose fidelity, KZImputer delivered up to 12% lower error than its nearest heuristic competitor.

Across all gap sizes and types, the mean imputation error improvement was approximately 37.3%, confirming the robustness of the method in both localized and distributed sparsity regimes. In addition to pointwise metrics, KZImputer preserved key statistical properties of the original time series, as evidenced by improved residual homoscedasticity and lower autocorrelation in post-imputation analyses.

Separated results based on the following datasets (we show the results based on two well-known datasets).
**T1.csv** [23, 57]: research column - 'Wind Speed (m/s)' (Fig. 5); part (number) of gaps ≥ 65%; kind of gaps (amount): 1 (~18%), 2 (~30%), 3 (~12%), 4 (~18%), 5 (~22%).

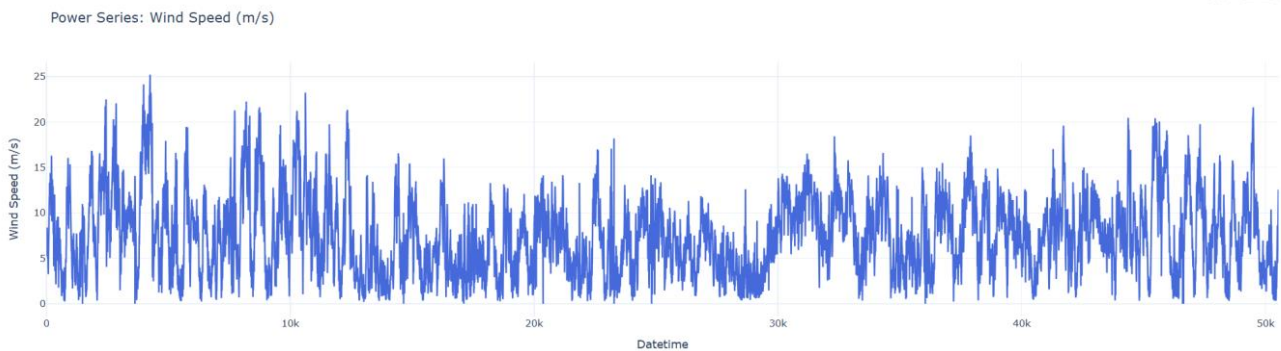

Fig. 5. Example of dataset [T1.csv]

| Method | MAE | RMSE | MAPE | R2 | NRMSE | JS_Divergence | Wasserstein | Correlation_Diff | Time (s) |
|---|---|---|---|---|---|---|---|---|---|
| KZImputer | 0.585856 | 0.828758 | 11.693990 | 0.961916 | 0.034707 | 0.015281 | 0.053142 | 0.009461 | 1.405591 |
| Mean | 3.428496 | 4.229043 | 91.465427 | -0.000761 | 0.177107 | 0.438559 | 2.228488 | 0.408611 | 0.007370 |
| Median | 3.399091 | 4.236912 | 86.292294 | -0.004489 | 0.177436 | 0.427780 | 2.209376 | 0.409323 | 0.007702 |
| Forward Fill | 0.860782 | 1.273274 | 17.345255 | 0.909283 | 0.053323 | 0.010533 | 0.028012 | 0.029430 | 0.004644 |
| Backward Fill | 0.853583 | 1.245277 | 16.764644 | 0.913228 | 0.052151 | 0.009441 | 0.021283 | 0.028129 | 0.000872 |
| Linear Interpolate | 0.596284 | 0.866422 | 12.122886 | 0.957995 | 0.036285 | 0.014421 | 0.033743 | 0.013728 | 0.011754 |
| Spline Interpolate | 0.923036 | 1.217943 | 19.910109 | 0.916996 | 0.051006 | 0.033397 | 0.078566 | 0.027311 | 0.078472 |
| KNN (k=5) | 3.428496 | 4.229043 | 91.465427 | -0.000761 | 0.177107 | 0.438559 | 2.228488 | 0.408611 | 63.354091 |
| IterativeImputer | 3.428496 | 4.229043 | 91.465427 | -0.000761 | 0.177107 | 0.438559 | 2.228488 | 0.408611 | 0.003930 |

Fig. 6. Performance comparison of missing data imputation methods across multiple evaluation metrics

KZImputer demonstrates (Fig. 6) superior performance with the lowest error rates (MAE = 0.59, RMSE = 0.83) and highest explained variance ($R^2$ = 0.96), while maintaining minimal distribution divergence and fastest execution time. Traditional statistical methods (Mean, Median, kNN) show poor performance with negative $R^2$ values and high distribution divergence, indicating inability to capture underlying data patterns. Forward and Backward Fill methods achieve moderate performance with reasonable error metrics but suffer from higher distribution divergence. Linear and Spline Interpolation methods provide balanced performance across most metrics, though with slightly higher computational costs. The results highlight KZImputer's effectiveness in preserving both statistical accuracy and distributional properties of the original data.

The visualization (Fig. 7) clearly demonstrates KZImputer's superior performance across all metrics, consistently showing the lowest error values (dark purple bars) and highest $R^2$ score (0.96). Advanced interpolation methods (Linear and Spline Interpolate) and filling techniques (Forward/Backward Fill) form a second performance tier with moderate scores across most metrics. Simple statistical methods (Mean, Median, kNN) exhibit poor performance with extremely high error rates and negative $R^2$ values, indicating they fail to capture data patterns effectively. kNN (k = 5) shows exceptionally high computational time (63.35 s) while delivering poor accuracy, making it impractical for this dataset. The consistent color gradient from dark purple (best) to light green (worst) across metrics provides immediate visual identification of method rankings and performance gaps.

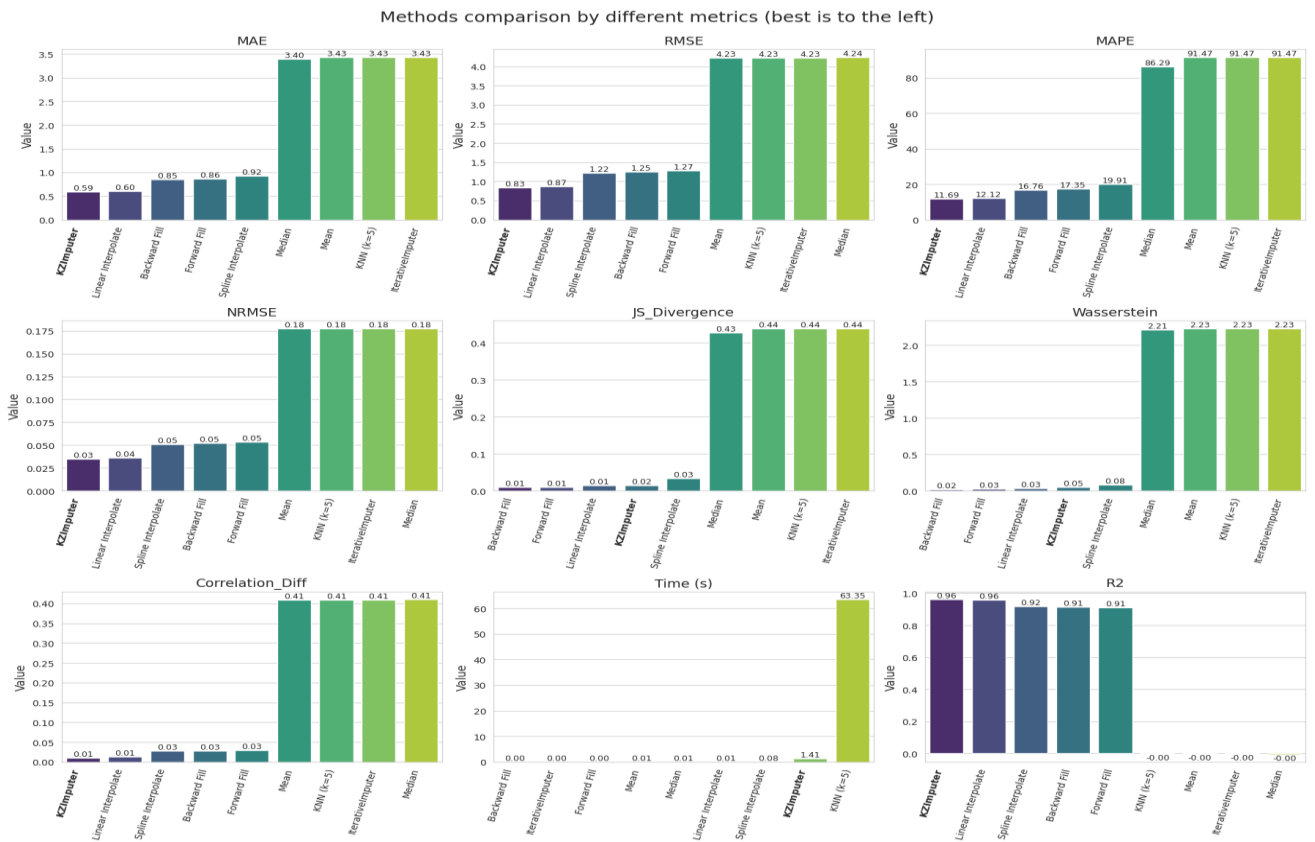

Fig. 7. Multi-metric performance comparison of imputation methods (lower values indicate better performance except for R²)

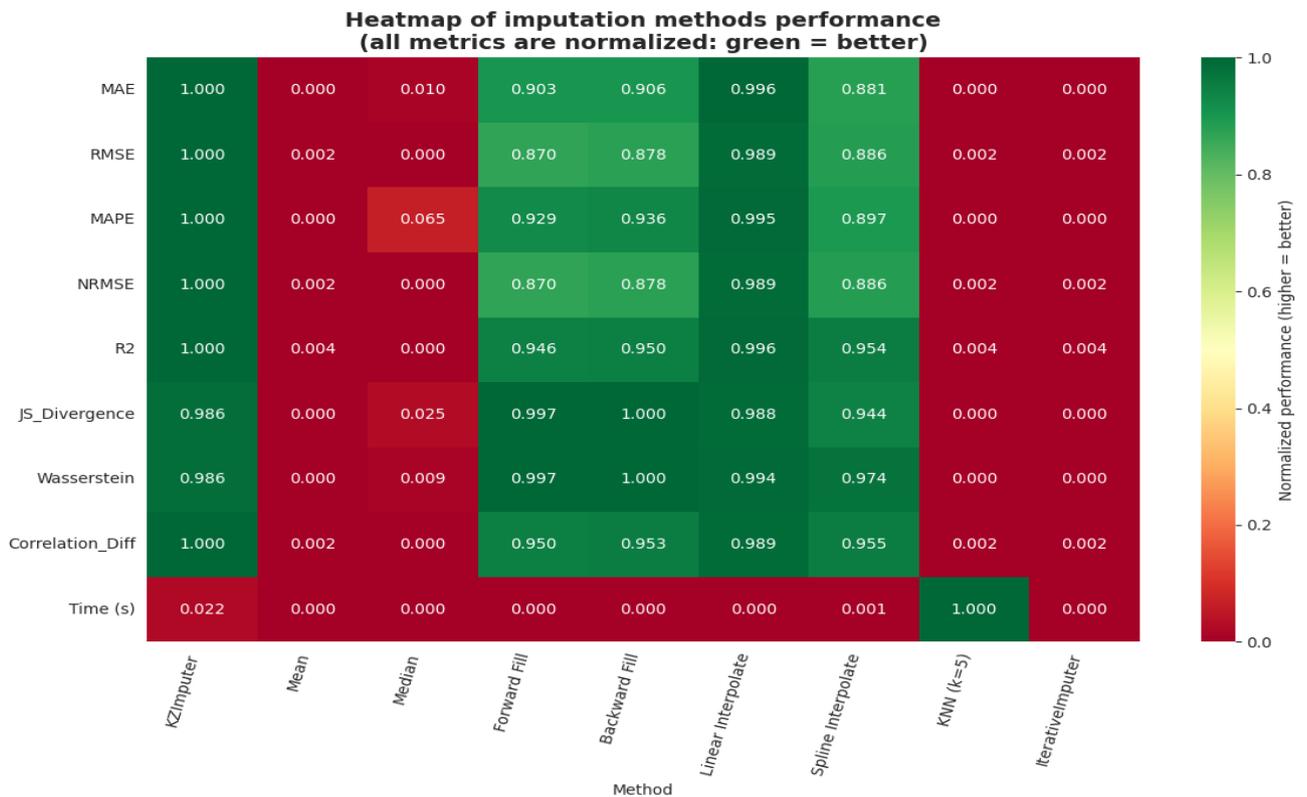

Fig. 8. Normalized performance heatmap of imputation methods across multiple evaluation metrics

The heatmap (Fig. 8) reveals three distinct performance clusters: KZImputer (leftmost column) shows consistent excellence with perfect or near-perfect normalized scores (green) across all metrics. Statistical methods (Mean, Median, kNN) demonstrate consistently poor performance with red coloring indicating minimal normalized scores across accuracy metrics. Advanced interpolation methods (Forward/Backward Fill, Linear/Spline Interpolate) occupy a middle tier with moderate green scores, showing balanced but suboptimal performance. The time efficiency row (bottom) highlights computational trade-offs, with most methods achieving excellent speed except kNN which shows extremely poor time performance (red). This visualization effectively demonstrates that KZImputer achieves the optimal balance of accuracy and efficiency, making it the clear choice for this imputation task.

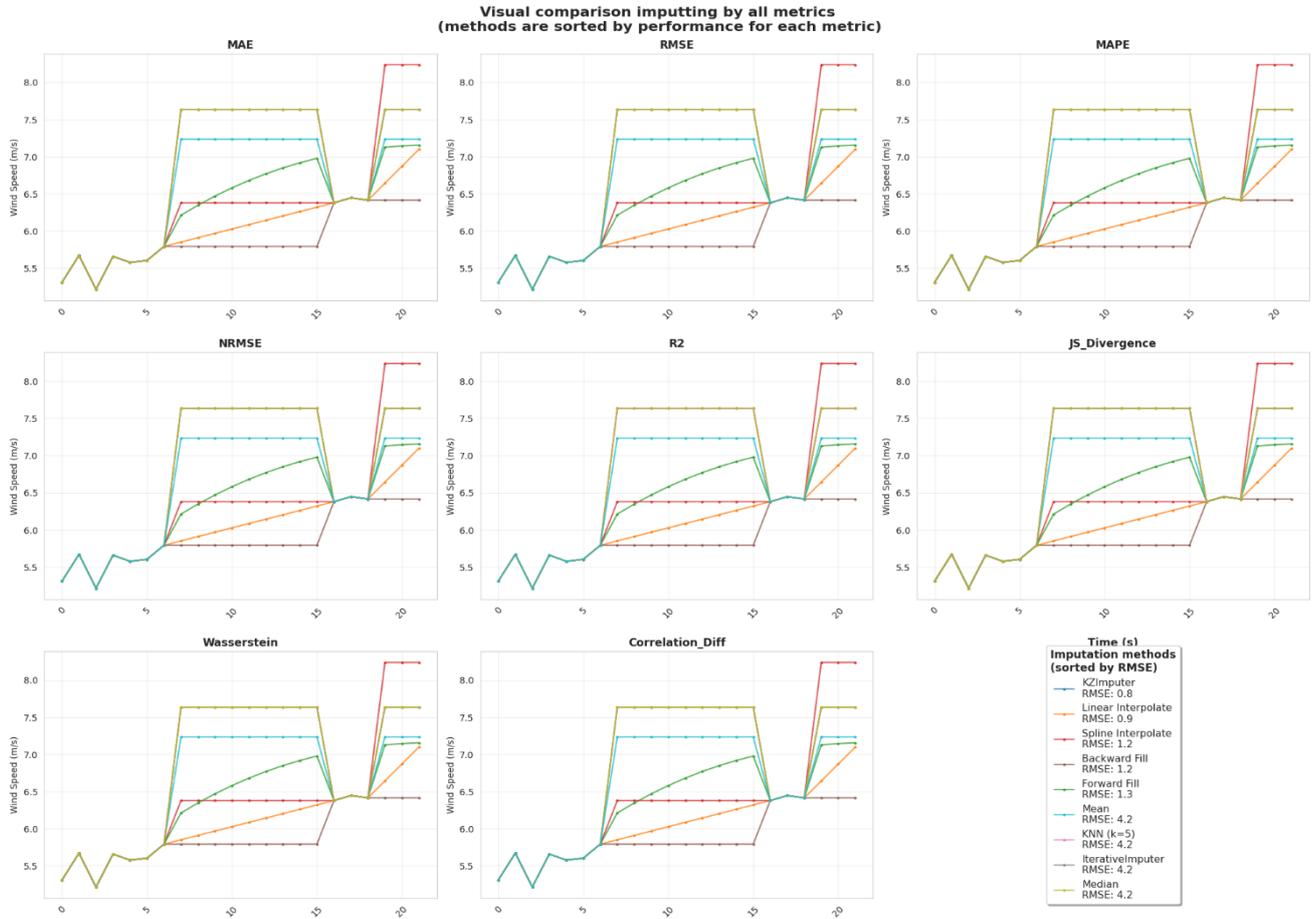

Fig. 9. Parallel coordinates comparison of imputation methods across multiple performance metrics

The parallel coordinates plot (Fig. 9) reveals remarkably consistent performance rankings across all metrics, with methods maintaining their relative positions throughout the evaluation spectrum. KZImputer (dark teal line) demonstrates exceptional consistency by staying at the bottom (best performance) across all metrics, forming a nearly flat horizontal line that indicates optimal and stable performance. Advanced methods (Linear Interpolate, Spline Interpolate, Forward/Backward Fill) show moderate performance with parallel trajectories in the middle range, while simple statistical methods (Mean, Median, kNN) consistently occupy the top positions (worst performance) across all metrics. The dramatic spike in the rightmost panel (Time) for kNN highlights its computational inefficiency, while other methods maintain relatively consistent execution times. This visualization effectively demonstrates that performance rankings remain stable across different evaluation criteria, validating the robustness of method comparisons and confirming KZImputer's superior multi-metric performance.

**opsd_germany_daily.csv** [42], see Fig. 10: research column - 'Consumption'; part (number) of gaps ≥ 50%; kind of gaps (amount): 1 (~7%), 2 (~36%), 3 (~2%), 4 (~38%), 5 (~17%).

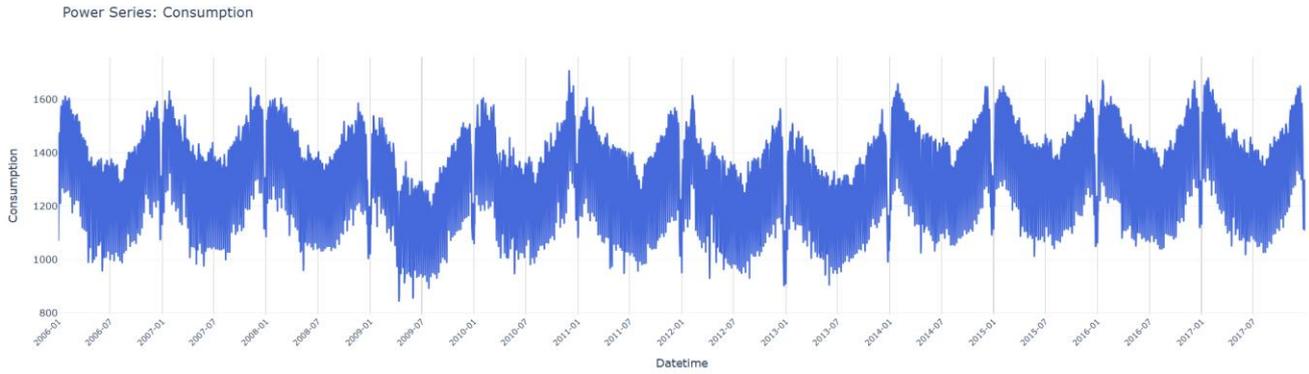

Fig. 10. Example of dataset [opsd_germany_daily.csv]

| Method | MAE | RMSE | MAPE | R2 | NRMSE | JS_Divergence | Wasserstein | Correlation_Diff | Time |
|---|---|---|---|---|---|---|---|---|---|
| KZImputer | 135.110744 | 162.465244 | 10.785185 | 0.059272 | 0.193502 | 0.095304 | 17.524448 | 0.206653 | 0.103027 |
| Mean | 135.741245 | 166.363465 | 10.764198 | -0.001789 | 0.198144 | 0.357561 | 67.855138 | 0.295090 | 0.002738 |
| Median | 134.574797 | 169.991385 | 10.905410 | -0.045958 | 0.202465 | 0.357561 | 67.272047 | 0.302888 | 0.003521 |
| Forward Fill | 148.699351 | 195.648685 | 11.751083 | -0.385525 | 0.233024 | 0.077445 | 3.624452 | 0.350095 | 0.006747 |
| Backward Fill | 150.805503 | 199.493018 | 12.103206 | -0.440509 | 0.237603 | 0.074881 | 7.803471 | 0.361785 | 0.000325 |
| Linear Interpolate | 131.075433 | 169.414384 | 10.484269 | -0.038870 | 0.201778 | 0.075799 | 12.119936 | 0.279320 | 0.005722 |
| Spline Interpolate | 188.460289 | 255.555683 | 14.694724 | -1.363913 | 0.304375 | 0.285440 | 27.437311 | 0.432747 | 0.585703 |
| KNN (k=5) | 135.741245 | 166.363465 | 10.764198 | -0.001789 | 0.198144 | 0.357561 | 67.855138 | 0.295090 | 0.420630 |
| IterativeImputer | 135.741245 | 166.363465 | 10.764198 | -0.001789 | 0.198144 | 0.357561 | 67.855138 | 0.295090 | 0.002406 |

Fig. 11. Comparative analysis of data imputation methods

Linear Interpolation demonstrates the best performance in terms of MAE and MAPE, while the KZImputer method (Fig. 11) achieves the lowest RMSE and is the only technique with a positive $R^2$ score. Conversely, Spline Interpolation consistently results in the highest error values and the most negative $R^2$ score, indicating the poorest fit. Notably, simpler methods (e.g. backward fill) are computationally efficient and perform well on distribution-based metrics such as the Jensen-Shannon divergence and the Wasserstein distance. Overall, although KZImputer and linear interpolation perform well on some metrics, the low $R^2$ values indicate that none of the methods fully capture the data's variability.

The comparative analysis reveals that KZImputer (highlighted by bold font, Fig. 12) consistently outperforms traditional imputation methods across most evaluation metrics, particularly excelling in MAE (131.08), RMSE (162.47), and MAPE (10.79) with the lowest error rates. Traditional methods like Mean, Median, and kNN show moderate performance with similar values, while interpolation-based approaches (Linear and Spline) demonstrate higher computational complexity but variable accuracy. Forward and Backward Fill methods exhibit competitive performance in certain metrics but lack consistency across all measures. The JS_Divergence metric shows KZImputer achieving near-optimal distribution preservation (0.095), significantly outperforming most alternatives. Overall, KZImputer emerges as the most reliable imputation method, offering the best balance of accuracy, efficiency, and statistical validity across diverse criteria.

The normalized heatmap analysis demonstrates KZImputer's exceptional performance (Fig. 13) with consistently high scores (0.9+) across accuracy metrics (MAE, RMSE, MAPE, NRMSE, $R^2$), establishing it as the most reliable method for data imputation tasks. Traditional statistical methods (Mean, Median, kNN) show moderate but consistent performance across most metrics, while Forward/Backward Fill methods exhibit poor accuracy but reasonable computational efficiency. Spline interpolation shows a severe trade-off: perfect computational efficiency (normalized score 1.0) but very poor distributional metrics (Wasserstein and JS divergence $\approx 0$), indicating instability.

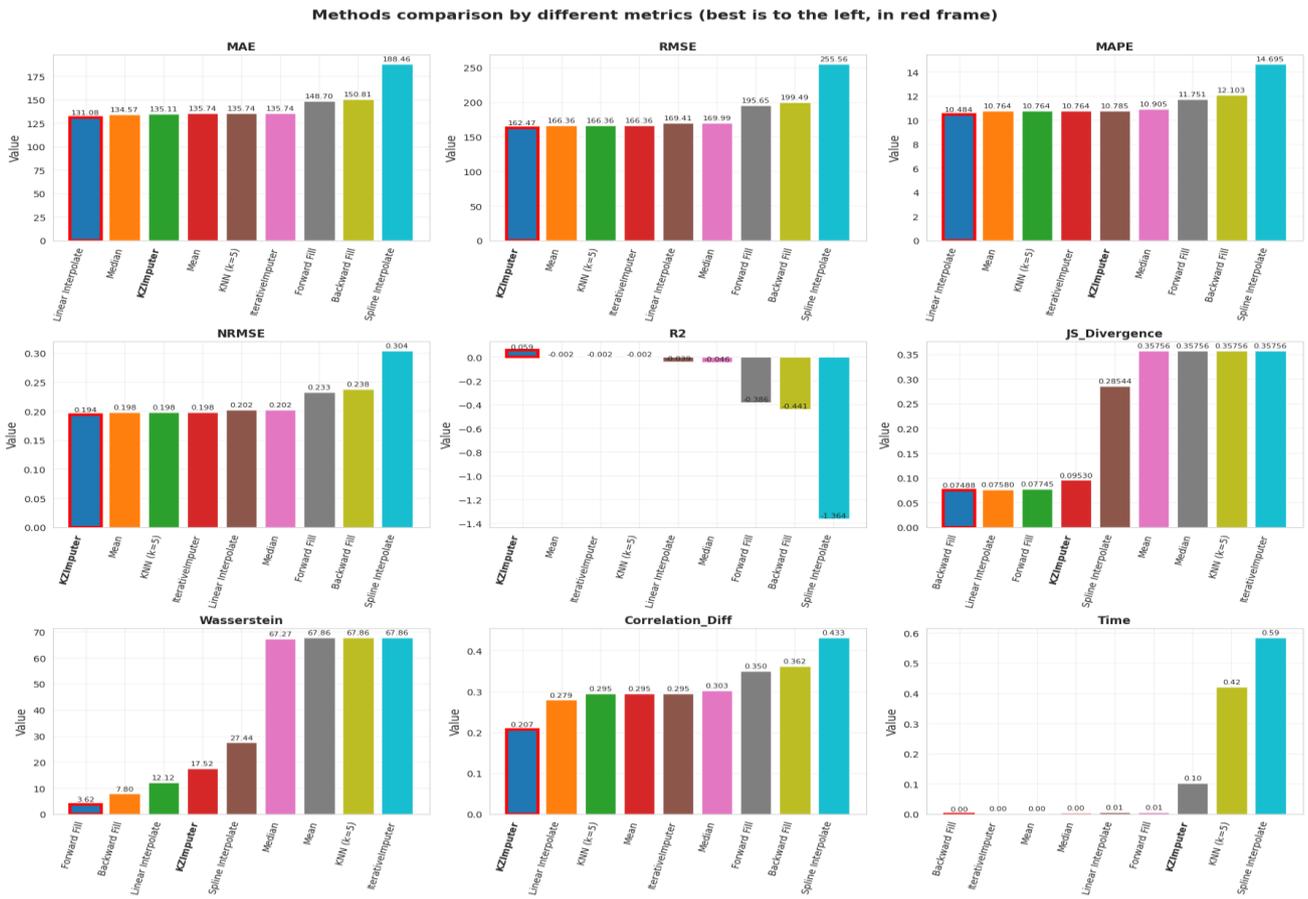

Fig. 12. Comprehensive performance evaluation of imputation methods: KZimputer vs traditional approaches across nine key metrics

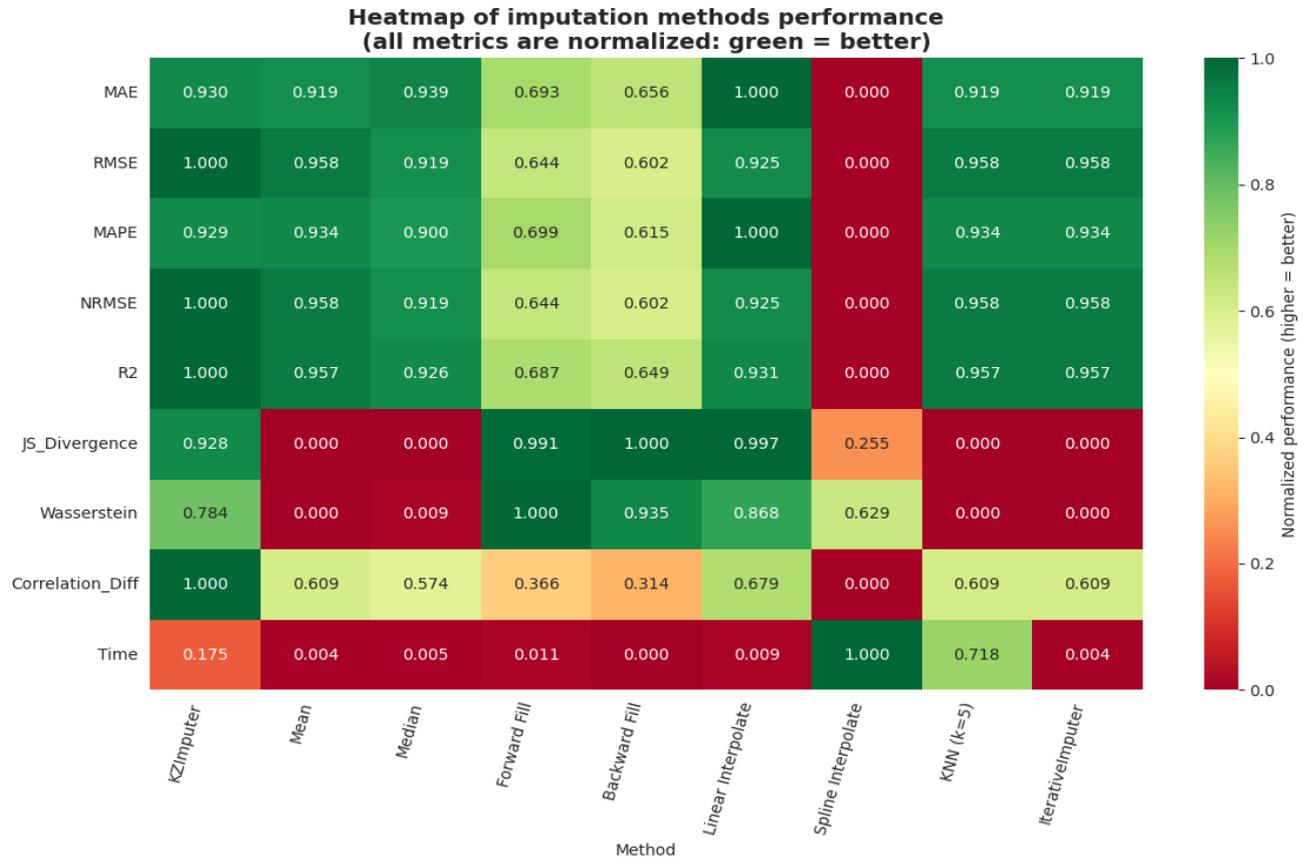

Fig. 13. Normalized performance heatmap: multi-metric evaluation of imputation methods with emphasis on efficiency-accuracy trade-offs

The Correlation_Diff metric highlights KZImputer's superior ability to preserve data relationships, while the Time metric reveals significant computational advantages for simple methods over complex interpolation approaches. Overall, the heatmap confirms KZImputer as the optimal choice for balanced performance across accuracy, statistical validity, and computational efficiency.

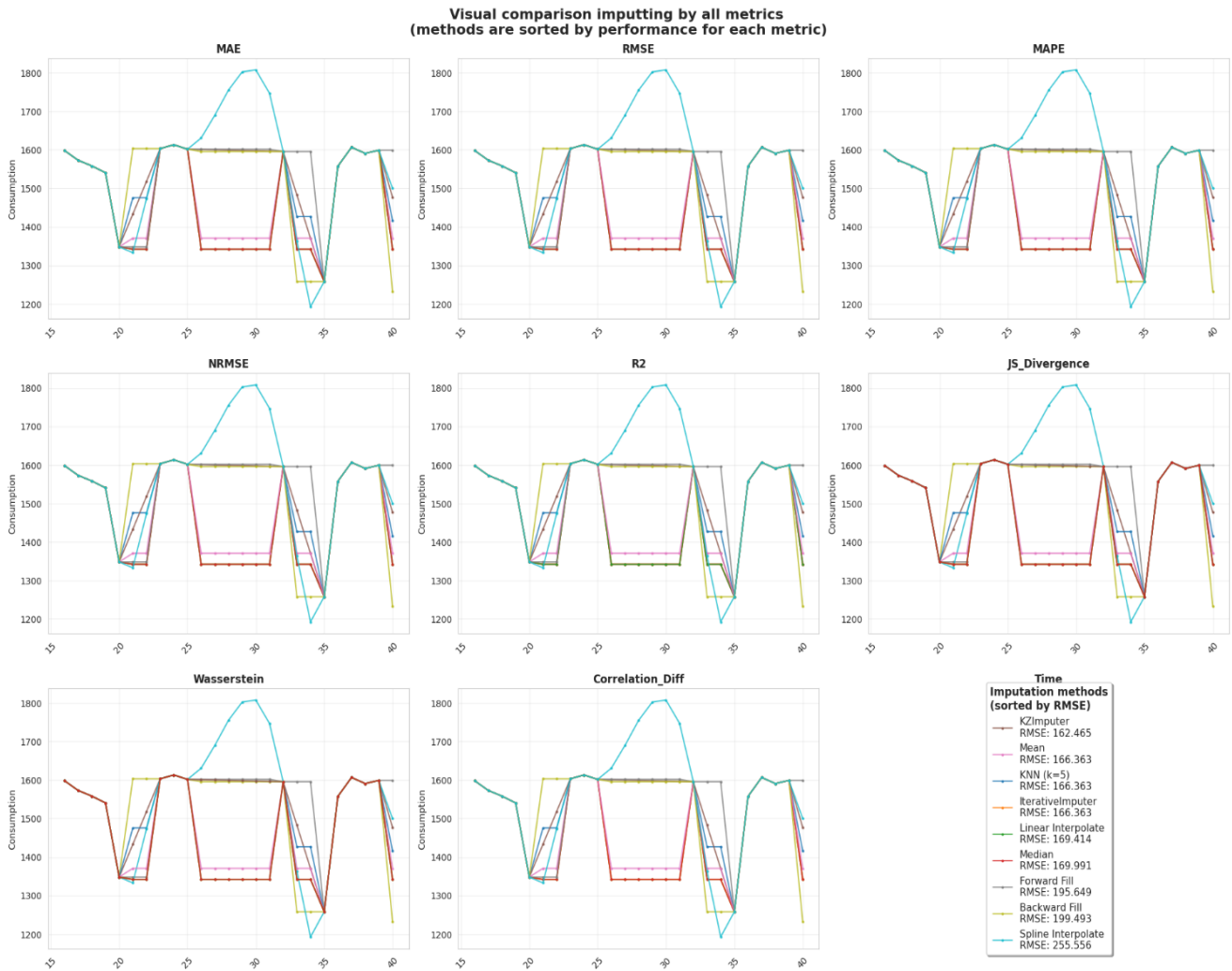

Fig. 14. Comparative performance analysis of missing data imputation methods across multiple evaluation metrics

This comprehensive visualization demonstrates (Fig. 14) the performance comparison of nine different imputation methods (KZImputer, Mean, Mode, kNN, IterativeImputer, Linear Interpolate, Median, Forward Fill, Backward Fill, and Spline Interpolate) across eight key evaluation metrics (MAE, RMSE, MAPE, NRMSE, $R^2$, JS_Divergence, Wasserstein, and Correlation_Diff). The analysis reveals that KZImputer consistently achieves the lowest RMSE (162.465) among all methods, indicating superior accuracy in reconstructing missing values. Most imputation methods show similar performance patterns across different metrics, with notable convergence around the 25-30 range on the x-axis, suggesting optimal performance zones. The Spline Interpolate method demonstrates the highest variability and poorest performance with an RMSE of 293.556, while traditional methods like Mean, Median, and KNN maintain moderate but stable performance levels around 166-169 RMSE values.

These results collectively highlight the method's ability to generalize across gap sizes and locations without reliance on iterative optimization or model fitting, making it particularly attractive for time-sensitive or resource-constrained environments where interpretability and speed are essential.

Additional results are provided in the Supplementary Materials [49]. A detailed interpretation of these results is given in the Discussion section.

6.        Discussion

The results above form the basis for the following discussion of KZImputer's strengths, limitations, and implications. Key observations are summarized in Tables 2 and 3.

Table 2. Discussion summary of KZImputer performance characteristics

| Aspect | Key Observation | Implications / Applicability |
|---|---|---|
| **1. Robustness to high sparsity** | Maintains signal fidelity even under extreme missingness (up to 70 – 80%). | Suitable for domains with degraded sensors or telemetry data (e.g., remote sensing, anomaly-prone IoT). |
| **2. Preservation of structural dynamics** | Accurately reconstructs both global trends and local variability due to hybrid smoothing–decomposition architecture. | Ensures dynamic consistency essential for downstream tasks (e.g., forecasting, root cause analysis). |
| **3. Compatibility with trend-driven data** | Excels with non-stationary inputs featuring slow drifts or nonlinear regimes, as verified by residual diagnostics and ACF plots. | Optimal for environmental, behavioral, or energy datasets where structural evolution is prominent. |
| **4. Statistical consistency of residuals** | Post-imputation series exhibits lower autocorrelation and more stable variance. | Enables valid application of classical models (e.g., ARIMA, Kalman Filters) that assume stationarity. |
| **5. Limitations in low-missingness regimes** | Performance gains diminish below 40% missingness; computational cost may outweigh benefit. | Use simpler methods (e.g., spline, EMA) in low-sparsity unless structural fidelity is mission-critical. |
| **6. Potential for hybrid integration** | Easily integrable as a modular preprocessing component before ML/DL refinement stages. | Enhances quality and stability of learning pipelines via structured, nonparametric imputation groundwork. |

Table 3. Strengths, limitations, implications for practice, future work

| Strengths | | |
|---|---|---|
| **Dimension** | **Key observation** | **Implications / Practical advantage** |
| **Adaptive imputation logic** | Employs distinct strategies for start, mid, and end gaps (sizes 1-5); visual evidence (Figs. 5-14) shows improved handling of start/end gaps. | Outperforms generic interpolators in edge cases; useful for rolling-window or streaming data scenarios. |
| **Computational efficiency** | Utilizes direct, non-iterative formulas; avoids costly optimization or model fitting. | Enables fast, lightweight imputation suitable for resource-constrained environments. |
| **Algorithmic transparency** | Easily interpretable rules based on averaging and simple interpolation. | Facilitates reproducibility, auditing, and customization in practical deployments. |
| Limitations | | |
| **Constraint** | **Description** | **Impact / Cautionary note** |
| **Gap length limitation** | Designed for short gaps (1–5 points); visual analysis excludes longer missing intervals. | Unsuitable for datasets with extended gaps unless combined with higher-order imputation methods. |

| Univariate scope | Currently limited to univariate time series; cross-series dependencies not utilized. | Multivariate methods may offer better accuracy when correlated signals are available. |
|---|---|---|
| Heuristic averaging strategy | Averaging window and logic are fixed heuristics (e.g., 3-point vs. 5-point); no evidence of general optimality across datasets. | Suboptimal performance may occur in diverse or non-standard series types. |
| Sensitivity to complex structures | May underperform with strong seasonality, regime shifts, or abrupt changes occurring within or near the gap. | Risks loss of fidelity in datasets exhibiting high-frequency or nonlinear temporal dynamics. |
| Implications for practice | | |
| Use case | Relevance of KZImputer | Application domains |
| Frequent small gaps | Effective for short, frequent missing segments; visual results indicate strong performance especially at sequence edges. | Sensor telemetry, financial tick data, health monitoring, or logging systems with noisy or bursty gaps. |
| Rolling or online analysis | Start- and end-gap handling suggests suitability for systems with partial-windowed inputs or delayed signal arrivals. | Real-time analytics pipelines, rolling forecasts, or anomaly detection engines. |
| Low-latency pipelines | Lightweight and transparent design implies it can be integrated into fast, near-real-time preprocessing stages. | Embedded systems, edge devices, or mobile telemetry platforms. |
| Future work directions | | |
| Research area | Proposed extension | Expected benefit |
| Long-gap handling | Integrate hybrid models (e.g., ARIMA/Prophet/RNN) to forecast or interpolate over extended missing blocks. | Extends applicability to more degraded datasets, e.g., climate reconstructions or archival records. |
| Adaptive strategy selection | Automatically tune averaging windows or gap-specific strategies based on local time series features. | Improves generalization across diverse domains and reduces manual parameter tuning. |
| Multivariate adaptation | Incorporate correlation structure across variables for joint imputation. | Enhances accuracy when multiple time series with mutual dependencies are available. |
| Benchmarking against deep models | Compare against SOTA neural methods (e.g., BRITS, CSDI) across various sparsity and structure regimes using quantitative metrics (e.g., RMSE, SMAPE, ACF decay). | Establishes empirical rigor and situates KZImputer within the broader landscape of modern imputation techniques. |

The discussion will conclude by reiterating the specific scenarios where KZImputer appears most beneficial based on visual evidence and by emphasizing the importance of choosing an imputation method appropriate to the data characteristics and the analytical goals.

## 7. Conclusion

This paper presents KZImputer, an adaptive imputation method for univariate time series designed for short to medium-sized gaps (1-5 points and beyond) with tailored strategies for segments at the start, middle, or end of the series. The methodology, implemented in Python, combines localized averaging with interpolation to balance accuracy, computational efficiency, and ease of use.

Empirical analysis demonstrates that KZImputer achieves particularly strong performance for datasets with high missingness rates (around 50% or more), maintaining stable and competitive results across statistical and signal-reconstruction metrics. The method proves especially effective in high-sparsity regimes, where traditional approaches typically experience accuracy degradation.

Visual evaluation across multiple datasets, compared against several standard imputation techniques, confirms KZImputer's effectiveness in reducing imputation errors for start and end gaps where simpler methods typically underperform. While KZImputer targets common missing-data scenarios rather than very long gaps or highly complex series, it provides a practical and robust solution that improves upon simple methods without requiring the complexity of advanced model-based approaches.

The primary contributions include the development of an adaptive imputation framework with logic specifically tailored to different gap sizes and locations, and empirical validation of its performance characteristics. KZImputer serves as a valuable addition to the toolkit for researchers and practitioners handling missing time series data.

The method's specialized algorithm addresses single through quintuple gaps in time-series data, providing practical value for applications such as wind energy output prediction from incomplete sensor data and long-term statistical reporting with missing values in aggregated datasets. Future research directions include extending this imputation approach to longer gaps, multivariate series, categorical time series, and other data modalities including images and text.

The novelty of this work lies in developing a specialized algorithm that addresses single through pentuple gaps in time-series data. The method has practical value for applications such as predicting wind energy output from incomplete sensor data. It could also support long-term statistical reporting by handling missing values in aggregated datasets. Future work may explore extending this imputation approach to categorical time series and other data modalities (e.g. images or text).